\documentclass[12pt]{article}

\usepackage[dvipdfm]{graphicx}
\begin{document}

\title{ Phase transition and   information cascade \\
in \\
a voting model }

\author{M Hisakado\footnote{[1]
masato\_hisakado@standardandpoors.com}  \space{} and   
S Mori\footnote{[2] mori@sci.kitasato-u.ac.jp} }

\maketitle

\begin{center}

*Standard  and Poor's, Marunouchi 1-6-5, Chiyoda-ku, Tokyo 100-0005, Japan

\vspace*{1cm}

\dag Department of Physics, School of Science,
Kitasato University, Kitasato 1-15-1 \\ 
Sagamihara, Kanagawa 252-0373, Japan

\end{center}

\vspace*{5cm}

\begin{abstract}
We introduce a voting model that is similar to a  
Keynesian beauty contest and analyze it from a mathematical point of view.  
There are two types of voters-copycat and independent-and two candidates. 
Our  voting model is a binomial distribution (independent voters)  doped in a 
beta binomial distribution (copycat voters). We find that  the phase 
transition in this system is at  the upper limit of  $t$,  where $t$ is the 
time (or the number of the votes). Our  model contains three phases. 
If copycats constitute a   majority or even  half of the total voters,
the voting rate  converges  more  slowly   than  it would  in a  binomial 
distribution. If independents constitute the majority of voters,
the voting rate converges   at the same rate  as it would  in a  binomial 
distribution. We also  study why it is difficult to estimate the conclusion 
of  a  Keynesian beauty contest  when there is  an  information cascade.
\end{abstract}



\newpage
\section{Introduction}

A Keynesian beauty contest is a popular concept used
to explain price fluctuations in equity markets.\cite{Keynes} 
Keynes described the action of rational agents in a market using 
an analogy based on a fictional newspaper contest. In the contest,  
entrants are asked to choose a set of  the six most beautiful  faces 
from  among  photographs of different women. Those  entrants who would 
select  the most popular face would be  then eligible for a prize.
A naive strategy would be  to choose  the most beautiful  face according 
to  the opinion of the entrant. Entrants  are known as employing such a 
strategy  independent voters. A more sophisticated  entrant, aiming  to 
maximize his/her chances of winning a prize, would try to deduce  the 
majority's perception of beauty.  This implies that the entrant would  
make a selection  on the basis of  some inference from his/her  knowledge 
of public perception.
Such voters are known as  copycats.
To estimate public perception,  people observe the actions of other 
individuals; then, they  make a  choice similar to  that of  others. 
Because it is usually sensible to do what other people are doing, 
the phenomenon is assumed to be the result of  a rational choice. 
Nevertheless, this approach can sometimes  lead to arbitrary or even 
erroneous decisions. This phenomenon  is called an  information cascade. 
\cite{Bikhchandani}

Collective herding phenomena in general pose quite interesting problems 
in  statistical  physics. To name a few examples, anomalous fluctuations 
in the  financial market \cite{Cont},\cite{Egu} and  opinion dynamics 
\cite{Stau}  have been related  to percolation and random field Ising model. 
A recent agent-based model proposed  by Curty and Marsili \cite{Curty} 
focused on the limitations  that herding imposed on the efficiency of  
information aggregation. Specifically, it was shown that when the fraction 
of herders in a population of agents increases, the probability that 
herding produces the correct forecast (i.e. that individual information 
bits are correctly aggregated) undergoes a transition to  a  state in 
which either all herders  forecast  rightly or no herder does.

We can observe  super-diffusive behaviour in the sense that  variance $D(L)$
grows asymptotically faster than $L$ (where $L$ is the long memory) in 
several fields.\cite{Kanter},\cite{Schenkel},\cite{Hod},\cite{Usatenko},
\cite{Huillet} It is  characterized by  the variance  $D\sim L^\alpha$ 
when $\alpha >1$. When $\alpha=1$, the diffusion of the variance becomes  
a standard Brownian motion.
For example,  in the case of daily financial   data,    $L$ represents 
the time series of data.
The past  price  affects the present price, and the diffusion becomes 
faster than  Brownian motion.
Such phenomena
can be attributed to  long-range positive correlations.
We may   observe  dynamical phase transition (from normal 
to super-diffusive behaviour).\cite{Hod}
In such a phase transition, correlation plays  an important role.
Further our voting model shows a similar transition.
The herders make  long-range correlations and display super-diffusive 
behaviour.
Therefore, a majority of voters reach    the wrong conclusion.

In this paper, we discuss  a  voting model  
with two candidates $C_1$ and $C_2$.
As mentioned above, we set two types of voters-independent and copycat.
Independent voters' voting is  based on  their fundamental values;
on the other hand, 
copycat voters' voting is  based on the number of votes.
In our previous paper, we
investigated the case wherein all the  voters are copycats.\cite{Mori}
In such a case, the process is a P\'{o}lya process, and  the voting rate  
converges to
 a beta distribution in a large time limit.\cite{Hisakado}
Our present  model   exhibits  a scale-invariant  behaviour. 
This behaviour is observed  in the mixing of the binary candidates. 
Furthermore, the power law holds over the entire range in  a double 
scaling limit.
This paper is an extension
of our previous works.

Although our model is very simple,
it contains  three phases.
We believe that  it is as adequate  as the percolation and random field 
Ising models, 
and  that it  is useful for understanding   phase transition in several fields.
We discuss   two specific issues:
one is 
 the  distribution in votes  that appears for  a mixture of independent  
and copycat voters and  
  the other is 
the change in  the vote  distributions over time.
On the basis of these above mentioned points, we discuss  phase transition  
for information cascade.

The organization of this paper is as follows.
In section 2, we introduce our  voting model and
 define the two types of  voters-independent and copycat-mathematically.
In section 3,
we  calculate the distribution functions strictly 
 for the special cases-independent voters always vote for either of the 
two  candidates;
their behavior is not probabilistic.
Then, we  obtain a solution
that is  an extension of the  solution given in 
\cite{Kullmann};
in this case, 
there is  no phase transition.
In section 4, we discuss
more general cases.
We use a stochastic differential equation, the  Fokker-Planck equation, 
and a numerical simulation.
In this model,
we can observe 
 phase transition at   the ratio
of  copycats to  independents through the variance of the distributions.
There are  three phases. 
If copycats constitute a  majority or number  half of the total 
number of voters,
 the voting  rate   converges  more slowly   than it would  in  a 
binomial distribution. If independents constitute the   majority of voters,
the voting  converges  at the  same rate as it would in  a  binomial 
distribution.
This implies that the proportion
 of copycats
influences the results of the  voting. 
 The last section presents the conclusions.

\section{Model}

\begin{figure}[h]
\begin{center}
\includegraphics[width=120mm]{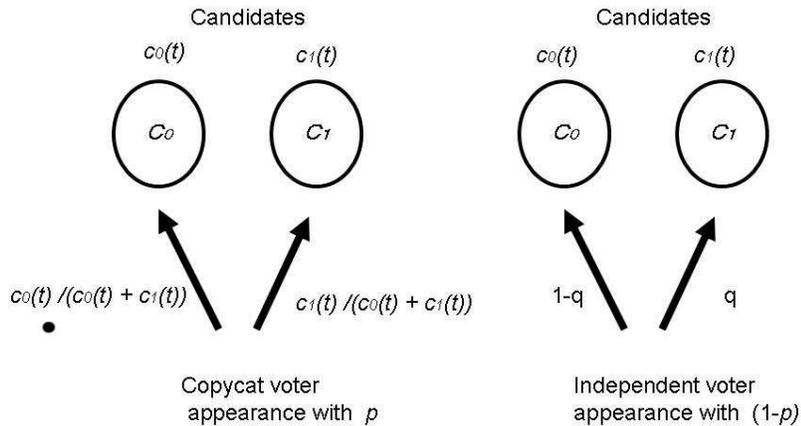}
\caption{Demonstration of  model.}
\end{center}
\end{figure}

We model the voting of two candidates, $C_0$ and $C_1$.
At  time $t$, each candidate has $c_0(t)$ and $c_1(t)$ votes.
At the beginning ($t=1$),
the two candidates, $C_0$ and $C_1$, have  $c_0(1)$ and $c_1(1)$ 
votes, respectively.
Hereafter,  we omit the time for the initial votes
 ($c_0\equiv c_0(1)$ and $c_1\equiv c_1(1)$)
and   define $c\equiv c_0+c_1$.
At each time step, one  voter  votes   for one  candidate. 
Voters are allowed see the number of votes for each candidate 
when they vote so that they have knowledge of   public perception.

There are  two types of voters-independent and copycat.
Independent voters vote for $C_0$ and $C_1$
with  probabilities $1-q$ and $q$, respectively.
Their votes are independent  of  others' vote
and  depend on what they think their fundamental value is.
Copycat voters vote for each
candidate 
with the probabilities  that are 
proportional  to   the candidates' votes.
If the number of votes 
are $c_0(t)$ and $c_1(t)$  at time  $t$,
a  copycat voter votes with  probability
$c_0(t)/(c_0(t)+c_1(t))$  for $C_0$ and $c_1(t)/(c_0(t)+c_1(t))$ for $C_1$.
Copycat voters' votes  are    based on the number of votes.

Here, we set the ratio of independent voters to 
copycat voters as $1-p$ and $p$, respectively.
If we set $p=1$, this system becomes 
a P\'{o}lya model  with 
$c_0$ starting elements of  type $C_0$  and  $c_1$ starting 
elements of  type $C_1$.
In this case, it is well known that the distribution 
of the  voting  rate  is a beta distribution.
As such, this system is a P\'{o}lya process doped  with a binomial distribution.

The evolution  equation for a candidate $C_0$ is
\begin{eqnarray}
P(k,t)&=&p\frac{k-1}{c_0+c_1+t-2}P(k-1,t-1)
+p(1-\frac{k}{c_0+c_1+t-2})P(k,t-1)
\nonumber \\
& &
+(1-p)qP(k,t-1)
+(1-p)(1-q)P(k-1,t-1).
\label{m1}
\end{eqnarray}
$P(k,t)$ is
the distribution of 
the number of votes $k$  at time $t$ for  candidate $C_0$.
The first and  second terms of (\ref{m1}) denote  the votes of  the
copycat voters;
the third and  fourth terms denote the votes of  the
independent voters.
If we set $Q(k,t)$ as the distribution of 
the number of votes $k$  at  time  $t$ for  candidate $C_1$,
we have the relation
\begin{equation}
Q(k,t)=1-P(c_0+c_1+t-1-k,t).
\label{re}
\end{equation}
The initial condition is
$
P(c_0,1)=1$.
This is the relation between the back and front.

\section{Exact solutions for $q = 1$ (or $q=0$)}

In this section, we study the exact solution of (\ref{m1}) for a special case.
For $q=1$, we obtain  the following master equation:
\begin{eqnarray}
P(k,t)&=&p\frac{k-1}{c+t-2}P(k-1,t-1)
+p(1-\frac{k}{c+t-2})P(k,t-1)
\nonumber \\
& &+(1-p)P(k,t-1).
\label{process}
\end{eqnarray}
At this limit, independent voters 
always vote for only one candidate, $C_1$
(if we set $q=0$, independent voters vote only for  $C_0$).
The master equation has a simpler form:
\begin{equation}
P(k,t)=P(k,t-1)-\frac{kp}{c+t-2}P(k,t-1)
+\frac{p}{c+t-2}(k-1)P(k-1,t-1).
\end{equation}
When we substitute $k=c_0$ in the above equation,   the last term of 
the RHS vanishes, 
and thus,  the probability $P(k,t)$ can   be calculated  easily:
\begin{equation}
P(c_0,t)=(1-\frac{c_0p}{c})(1-\frac{c_0p}{c+1})\cdots (1-\frac{c_0p}{c+t-2}).
\end{equation}
For $k>C_0$, we can prove (see Appendix A) that  the following equality holds:
\begin{eqnarray}
 P(k',t)
&=&
\sum_{l=1}^{k'-c_0+1}(-1)^{l-1}\frac{(c_0)_{k'-c_0}}{(1)_{k'-c_0-l+1}(1)_{l-1}}
\frac{(c-(l+c_0-1)p)_{t-1}}{(c)_{t-1}}.
\label{distribution}
\end{eqnarray}
This is  the distribution of the votes  for the special case wherein  
the independent voters always vote for only one candidate, $C_1$.

If  we set $p=1$,  all voters are copycats, and we obtain the following 
reduction:
\begin{eqnarray}
P(k',t)&=&
\frac{(c_0)_{k'-c_0}}{(c)_{t-1}(1)_{k'-c_0}}
\sum_{l=1}^{k'-c_0+1}(-1)^{l-1}
\left(
\begin{array}{cc}
k'-c_0&\\
l-1&\\
\end{array}
\right)
(c_1-l+1)_{t-1},
\nonumber \\
&=&
\left(
\begin{array}{cc}
t-1&\\
k'-c_0&\\
\end{array}
\right)
\frac{(c_0)_{k'-c_0} (c_1)_{t-1-k'+c_0}}{(c)_{t-1}}.
\label{13}
\end{eqnarray}
This is a beta binomial distribution.
At the limit $t\rightarrow \infty$, the above equation  becomes a 
beta distribution.
Note that to obtain (\ref{13}), we use the identity
\[
\sum_{l=1}^{k'-c_0+1}(-1)^{l-1}
\left(
\begin{array}{cc}
k'-c_0&\\
l-1&\\
\end{array}
\right)
(c_1-l+1)_{t-1}
=\frac{(1)_{t-1}}{(1)_{t-1-k'+c_0}}(c_1)_{t-1-k'+c_0}.
\]
In \cite{Mori}, we discussed the physical characteristic of this model. 
In the limit 
$t\rightarrow \infty$ and $c_0,c_1 \rightarrow 0$ with $\alpha=c_1/c_0$ 
fixed, the scale invariance holds over the entire range.

Here, we can  calculate the momentum of these distributions to  analyze them.
The momentum is given by
\begin{equation}
\mu_r(t)=\sum_{k=c_0}^{c_0+t-1}k^rP(k,t).
\end{equation}
We introduce  quasi-momentum as
\begin{equation}
\hat{\mu}_r(t)\equiv\sum_{k=c_0}^{c_0+t-1}k(k+1)\cdots(k+r-1)P(k,t).
\label{mu}
\end{equation}

We can prove (see Appendix B) that  the quasi-momentum can  have the 
following form:
\begin{eqnarray}
& &\hat{\mu}_r (t)=
\sum_{l=1}^{t}(-1)^{l-1} \frac{(c_0)_{t+r}}{(1)_{l-1}(1)_{t-l}(l+c_0+r-1)}
\frac{(c-(l+c_0-1)p)_{t-1}}{(c)_{t-1}}.
\nonumber \\
& &
\label{moment}
\end{eqnarray}
If we set $r=1$, we  get the
average vote
\begin{eqnarray}
& &\hat{\mu}_1 (t)=
\sum_{l=1}^{t}(-1)^{l-1} \frac{(c_0)_{t+1}}{(1)_{l-1}(1)_{t-l}(l+c_0)}
\frac{(c-(l+c_0-1)p)_{t-1}}{(c)_{t-1}}.
\end{eqnarray}

We study  $t\rightarrow \infty$.
The coefficients of master equation (\ref{process}) do not contain
the initial votes for $C_0$; given by  $c_0$.
If we set $t>>c=c_0+c_1$, the master equation does not depend on the
initial conditions.
Therefore, for a large $t$ limit, the behaviour of the moment does not 
depend on the initial conditions $c_0$, and $c_1$.
We can also  observe this in  (\ref{fp}) and (\ref{ito}) in the  next section.  
Here, we set $c_0=c_1=1$ for the representative case.
Direct calculation  using (\ref{moment}) is difficult.
Hence, we study $P(k',t)$ as 
$t\rightarrow \infty$.
In  the above case,  distribution (\ref{distribution}) becomes 
a constant distribution with  cut-off $k^*$.
The cut-off implies  a fast decay for larger values ($k'>k^*$).
If we set $p=1$, we get a constant distribution.
Using  (\ref{distribution}), we get
\begin{equation}
\lim_{t\rightarrow \infty}P(k',t)
=
\sum_{l=1}^{k'}(-1)^{l-1}\frac{(1)_{k'-1}}{(1)_{k'-l}(1)_{l-1}}
t^{-p}.
\label{limit}
\end{equation}
For large time values, the only time dependent term 
is $t^{-lp}$.
In the case of $t>>k'$, we can assume that  only the first term of 
the summation is non-negligible. 
Therefore,
\begin{equation}
\lim_{t\rightarrow \infty}P(k',t)
\sim
t^{-p}.
\end{equation}
Cut-off  $k^*$ is  the inflection point of $P(k',t)$.
Using (\ref{limit}), we can obtain $k^*=t^p+2$.
Then, the momentum is
\begin{equation}
\mu_r(t)\sim\sum_{k=1}^{k^*}k^rP(k,t)\sim t^{rp}.
\label{3}
\end{equation}
(\ref{3}) is a continuous function of $p$.
Hence, there is no phase transition throughout.
In the next section, we study the general case 
in the continuous limit.

\section{Asymptotic cases}
To investigate   long-ranged correlations, we 
analyze in the limit $t \rightarrow \infty$.  
We can rewrite  (\ref{process}) as
\begin{eqnarray}
c_0(t)=k \rightarrow k+1:
 P(k,t)&=&\frac{kp}{c+t-1}+(1-p)(1-q)
\nonumber \\
&=&\frac{p}{2}(1+\frac{2k-(c+t-1)}{c+t-1})+(1-p)(1-q).
\nonumber  \\
\label{pd}
\end{eqnarray}
We define 
$\Delta_t=2c_0(t)-(c+t-1)$ with the initial condition 
$\Delta_1=c_0-c_1=2c_0-c$. We change the notation from 
$k$ to $\Delta_t$ for convenience.
Then, we have $|\Delta_t|=|2k-(c+t-1)|<c+t-1$.
The support for the law of $\Delta_t$ is thus 
$\{\Delta_1-(t-1),\Delta_1+(t-1)\}$. Given $\Delta_t=s$, 
we obtain a random walk model
\begin{eqnarray}
\Delta_t&=&s \rightarrow s+1  :P_{\frac{s+c+t-1}{2},t}=\frac{p(s+c+t-1)}{2(c+t-1)}
+(1-p)(1-q),
\nonumber \\
\Delta_n&=&s \rightarrow s-1  :Q_{\frac{s+c+t-1}{2},t}=1-P_{\frac{s+c+t-1}{2},t}.
\nonumber
\end{eqnarray}
Let $\epsilon=1/c\rightarrow 0$. We now  consider
\begin{eqnarray}
X_\tau&=&\epsilon\Delta_{[t/\epsilon]},
\nonumber \\
P(x,\tau)&=&\epsilon P(\Delta_t/\epsilon,t/\epsilon),
\end{eqnarray}
where $\tau=t/\epsilon$ and $x=\Delta_t/\epsilon$.
Approaching the continuous limit, we can obtain the Fokker-Plank  diffusion 
equation for this process (see Appendix C):
\begin{equation}
\frac{\partial P}{\partial \tau}
=\frac{1}{2}
\frac{\partial^2 P}{\partial x^2}-
\frac{p}{\tau+1}\frac{\partial (xP)}{\partial x}
-(1-p)(1-2q)\frac{\partial P}{\partial x}.
\label{fp}
\end{equation}
We can also  obtain  $X_\tau$ such that it  obeys a diffusion 
equation with small additive noise:
\begin{equation}
\textrm{d}X_\tau=[(1-p)(1-2q)+\frac{px}{\tau+1}]\textrm{d}\tau+\sqrt{\epsilon}
\textrm{d}B_\tau,\hspace{2cm}
X_0=\frac{c_0-c_1}{c}.
\label{ito}
\end{equation}
Though (\ref{fp}) and (\ref{ito}) are equivalent, 
hereafter, we only deal with (\ref{ito}) for  simplicity. 
Assume $c_0$ is random or deterministic.
Let 
\begin{equation}
\sigma^2_0\equiv\sigma^2(Y_0)=4\epsilon^2\sigma^2(s)
\end{equation}
be the variance of $X_0$.
If $X_0$ is Gaussian   $(X_0\sim\textrm{c}(y_0,\sigma^2_0))$ 
or deterministic $(X_0\sim\delta_{x0})$, the law of $X_\tau$ 
ensures that the  Gaussian is in  accordance with density
\begin{equation}
p_\tau(x)\sim
\frac{1}{\sqrt{2\pi}\sigma_\tau}\textrm{e}^{-(x-x_\tau)^2/2\sigma_\tau^2},
\end{equation}
where $x_\tau=\textrm{E}(X_\tau)$ is the expected value of $X_\tau$ and 
$\sigma^2_\tau\equiv v_\tau$ is its variance.
If $\Phi_\tau(\lambda)=\log(\textrm{e}^{\textrm{i}\lambda X_\tau})$
 is the logarithm of the characteristic function of the 
law of $X_\tau$, we have\begin{equation}
\partial_\tau
\Phi_\tau(\lambda)
=\frac{p}{1+\tau}\lambda\partial_\lambda\Phi_\tau(\lambda)
+\textrm{i}(1-p)(1-2q)\lambda-\frac{\epsilon}{2}\lambda^2
\end{equation}
and
\begin{equation}
\Phi_\tau(\lambda)=\textrm{i}\lambda x_\tau-\frac{\lambda^2}{2}v_\tau.
\end{equation}
Identifying
the real and imaginary parts of $\Phi_\tau(\lambda)$, we 
obtain the dynamics of the mean  of $X_\tau$ as
\begin{equation}
\dot{x}_\tau=\frac{p}{1+\tau}x_\tau+(1-p)(1-2q).
\end{equation}
The solution for $x_\tau$ is
\begin{equation}
x_\tau=(x_0+2q-1)(1+\tau)^p+(1-2q)(1+\tau)\sim (x_0+2q-1)\tau^p+(1-2q)\tau.
\end{equation}
Since we are interested in the voting  rate obtained, 
 we introduce a new scaled variable:
\[
\tilde{x}_\tau\equiv \frac{x_\tau}{\tau}.
\]
The solution for $\tilde{x}_\tau$ is
\begin{equation}
\tilde{x}_\tau\sim(x_0+2q-1)\tau^{(p-1)}+(1-2q).
\end{equation}
When $p\neq1$, $\tilde{x}_\tau\sim 1-2q$.
This implies that  the average  percentage of $C_0$'s votes againt 
the total poll is
$1-q$.
When $p=1$, $\tilde{x}_\tau\sim x_0$.
This agrees with our assertion that the  scaled distribution of votes  
becomes a beta distribution when $\tau$ is large. 
In this case, the mean value does not change.

From the above discussion, we can infer  that the distribution becomes 
similar to  a delta function.
The question of how this  distribution converges to a delta function
constitutes  the next  problem.
To investigate this, we analyze the dynamics of the variance. 
The dynamics of $v_\tau$ are  given by the Riccati equation
\begin{equation}
\dot{v}_\tau=\frac{2p}{1+\tau}v_\tau+\epsilon.
\label{gp}
\end{equation}
If $p\neq 1/2$, we get
\begin{eqnarray}
v_\tau&=&
v_0+\int_0^\tau(\frac{1+\tau}{1+r})^{2p}(\frac{2p}{1+r}v_0+\epsilon)\textrm{d}r
=v_0(1+\tau)^{2p}+\epsilon\int_0^\tau(\frac{1+\tau}{1+r})^2p\textrm{d}r
\nonumber \\
&=&
v_0(1+\tau)^{2p}+\frac{\epsilon}{1-2p}(1+\tau)^{2p}((1+\tau)^{1-2p}-1).
\end{eqnarray}
If $p=1/2$, we get
\begin{equation}
v_\tau=v_0+\int_0^{\tau}(\frac{1+\tau}{1+r})(\frac{1}{1+r}v_0
+\epsilon)\textrm{d}r=v_0(1+\tau)+\epsilon(1+\tau)\textrm{log}(1+\tau).
\end{equation}
Now, we can  summarize the temporal behaviour of the variance as 
\begin{equation}
v_\tau\sim\frac{\epsilon}{1-2p}\tau\hspace{1cm}\textrm{if}
\hspace{0.5cm}p<\frac{1}{2},
\end{equation}
\begin{equation}
v_\tau\sim(v_0+\frac{\epsilon}{2p-1})\tau^{2p}\hspace{1cm}\textrm{if}
\hspace{0.5cm}p>\frac{1}{2},
\end{equation}
\begin{equation}
v_\tau\sim\epsilon\tau\textrm{log}(\tau)\hspace{1cm}\textrm{if}
\hspace{0.5cm}p=\frac{1}{2}.
\end{equation}
Here, we introduce rescaled variables
\[
\tilde{v}_\tau\equiv\frac{v_\tau}{\tau^2}.
\]
The solution for $\tilde{v}_\tau$ is
\begin{equation}
\tilde{v}_\tau\sim\frac{\epsilon}{1-2p}\tau^{-1}\hspace{1cm}\textrm{if}
\hspace{0.5cm}p<\frac{1}{2},
\end{equation}
\begin{equation}
\tilde{v}_\tau\sim(v_0+\frac{\epsilon}{2p-1})\tau^{2p-2}\hspace{1cm}
\textrm{if}\hspace{0.5cm}p>\frac{1}{2},
\end{equation}
\begin{equation}
\tilde{v}_\tau\sim\epsilon\frac{\textrm{log}(\tau)}{\tau}
\hspace{1cm}\textrm{if}\hspace{0.5cm}p=\frac{1}{2}.
\end{equation}
If $p=1$, $\tilde{v}_\tau$ becomes $v_0$. This agrees with our assertion 
that  the distribution of votes becomes a beta distribution.
If $p>1/2$ or $p=1/2$,
   candidate $C_0$ gathers $1-q$ of all the  votes in the scaled 
distributions, but  the voting rate   converges more slowly than that 
in  a binomial distribution.
If  $0<p<1/2$, the voting rate  becomes $1-q$, and  the distribution 
converges as  it would in a binomial distribution.
Hence, if independent voters form a majority,
the distribution of votes   becomes similar to  a delta function and 
the convergence is
 at the same rate as  that  in a binomial distribution.
If  copycat  voters form  the majority,
the distribution remains the same but the convergence is  at a rate slower
than that  in a binomial distribution.
In this phase,  it is difficult  to ascertain  the causes  for the 
delay of the convergence. Similar phenomena  can be seen in several fields.  
In  daily financial data, the motion of the price does not represent a  
Markov process, and it is difficult to forecast the future price.
\cite{Usatenko}   In fact, it has been  pointed out that the motion of 
price is super-diffusive behaviour and the stochastic differential 
equation for  price  is similar to (\ref{ito}).\cite{Hod}
When all voters are copycats,
the distribution becomes a beta distribution and does not converges.

Curty and Marsili \cite{Curty} recently introduced a model about 
information cascade. Their model  is based on  game theory. They showed  
that when the fraction of herders in a population of agents increases, 
the probability that herding produces the correct forecast undergoes a 
transition to  a  state in which either  all herders  forecast  
rightly or no herder does.   Their model is similar to the limitation  
of our model in the case wherein  voters  are unable to see  the votes 
of  all the voters but can only  see   the votes of previous voters. 
However, there is a  significant  difference between our model and their 
model with  respect to the  behaviour of copy cats.  In their model, 
copycats always  select the majority of votes, which  is visible to them.  
Thus,  the behaviour becomes      digital (discontinuous).  We aim to 
carry out an analysis   of the influence of this behaviour in the future.

We now consider the correlation. For a beta binominal distribution, 
we can define the parameter $\rho\equiv 1/(c+1)$.\cite{Hisakado} 
This parameter  represents the strength of following a decision.
If we set $\rho=1$, everyone votes for the candidate who received   
the first voter's vote. On the other hand,
when  $\rho=0$, copycats  become independent.
It should be noted  that our  conclusion does not depend 
on $\rho$ except when $\rho=0$.
For large $t$, convergence is not related to $\rho$, 
but is related  to  $p$, the appearance probability of independent voters.

 Here we discuss the solution in the previous section. 
If we set $q=0$ ($q=1$  is the same as  relation(\ref{re})), 
(\ref{pd}) becomes $0$ for  large $t$. 
This is  because   independent voters' votes become deterministic.
Hence, in (\ref{fp}), the diffusion term disappears.
In (\ref{gp}),  the noise term $\epsilon$ disappears, and 
the dynamics of   $v_\tau$ are  given by
\begin{equation}
\dot{v}_\tau=\frac{2p}{1+\tau}v_\tau.
\end{equation}
Then,  phase transition disappears, and the behaviour of  $v_0$ is continuous
\begin{equation}
\tilde{v}_\tau\sim(v_0)\tau^{2p-2}\hspace{1cm}\textrm{for all}\hspace{0.5cm}p.
\end{equation}
This result is acceptable, following  the discussion in the 
previous section and that (\ref{moment}) is  continuous  with respect to $p$.
(See (\ref{3}).)
If there is a consensus about the fundamental value,
copycat voters  affect the  convergence in proportion to their  
ratio. 
Further, in this case, 
the  convergence   does not depend on  correlation $\rho$.

In order to confirm the analytical results pertaining to the 
asymptotic behaviour,
we perform numerical  simulations.
We use the master equation (\ref{1}) directly.

Figures 2 and 3 display the deformation of the distribution of  
votes for $C_0$  over  time $t$. Figure 2  is  the case  wherein the 
independent voters vote for $C_0$ with the probability $1-q$, and 
Figure 3  is  the case wherein the  independent voters  always vote 
for $C_0$.
We can see that  the distribution converges to a delta function for 
the votes of the independent voters. If all voters are copycats ($p=1$), 
the distribution becomes a beta binomial distribution. Because of the   
doped binomial distribution (independent voters), the distribution 
is deformed. For the case $q=0$, we can obtain  an  
exact solution in section 3.   

\begin{figure}[h]
\begin{minipage}{.5\textwidth}
\includegraphics[width=\textwidth]{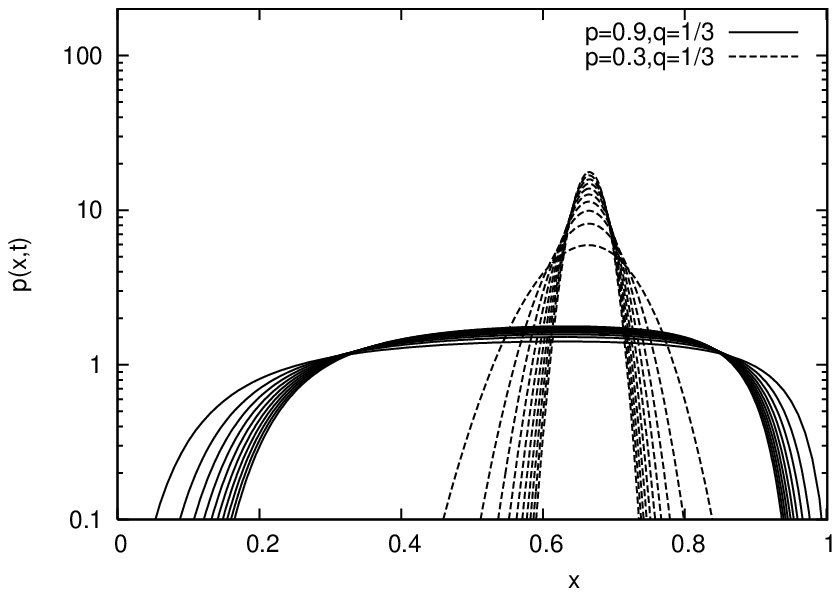}
\label{fig3}
\caption{Asymptotic behaviour of the distribution of 
votes for $C_0$ at $t=100 \rightarrow 10000$ for $q=1/3$ and $p=0.9,0.3$.}
\end{minipage}
\hspace{0.3cm}
\begin{minipage}{.5\textwidth}
\includegraphics[width=\textwidth]{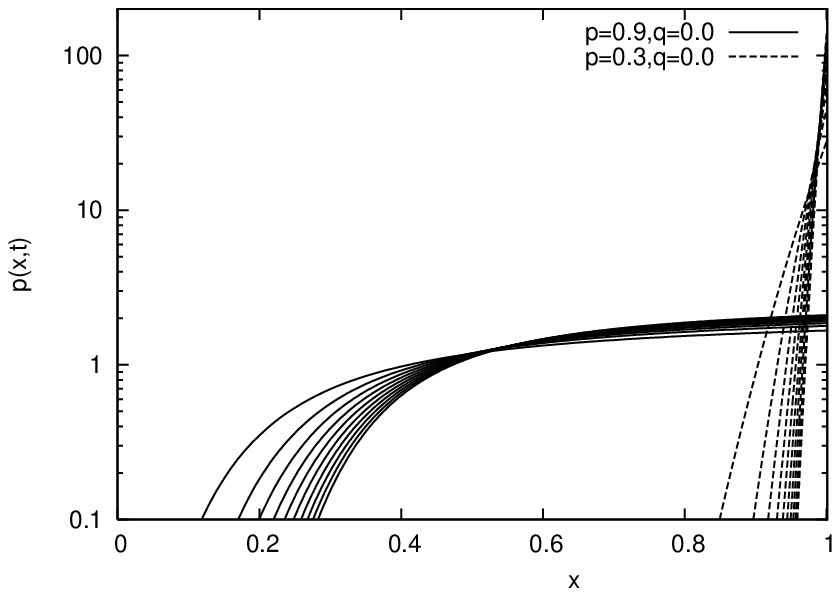}
\label{fig4}
\caption{Asymptotic behaviour of the distribution of votes 
for $C_0$  at $t=100 \rightarrow 10000$ for  $q=0$ and
$p=0.9,0.3$.}
\end{minipage}
\hspace{0.3cm}
\end{figure}

Figures  4 and  5  display the resulting scaled variance for 
 different  $p$.
The distribution converges  to a delta function  over   time. 
However, there are  differences between Figures 2 and  3 at $p\geq 0.5$.
The difference  is  in the speed of the convergence 
that  is characterized by the scaled variance $\tilde{v}$.
In the general case ($q\neq 0,1$),
we can recognize  the phase transition (Figure  4).
If $p>0.5$,  the variance converges at the same rate as that  in
a binomial distribution (slope $=-1$).
If $p<0.5$ or $p=0.5$,  super-diffusive behaviour is exhibited, and  
convergence is 
 slower than that   in a binomial distribution (slope $> -1$).
The cases $p<0.5$ and  $p=0.5$ represent two different phases.
\begin{figure}[h]
\begin{minipage}{.5\textwidth}
\includegraphics[width=\textwidth]{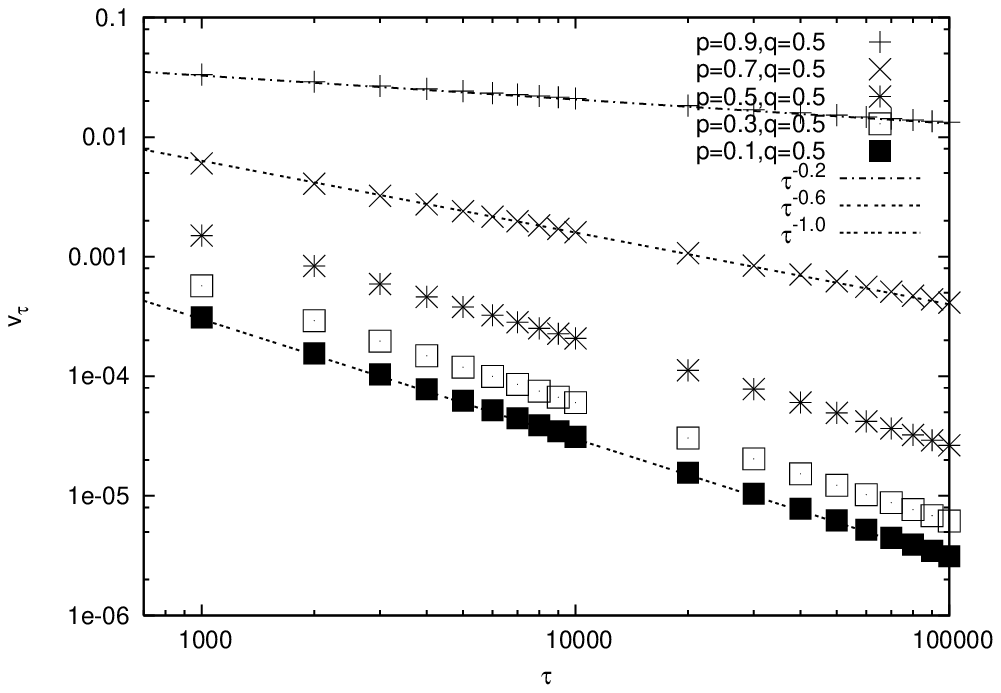}
\label{1}
\caption{Asymptotic behaviour of the scaled variance 
$\tilde{v}_\tau$ for $q=0.5$ and $p=0.9,0.7,0.5,0.3,0.1$.}
\end{minipage}
\hspace{0.3cm}
\begin{minipage}{.5\textwidth}
\includegraphics[width=\textwidth]{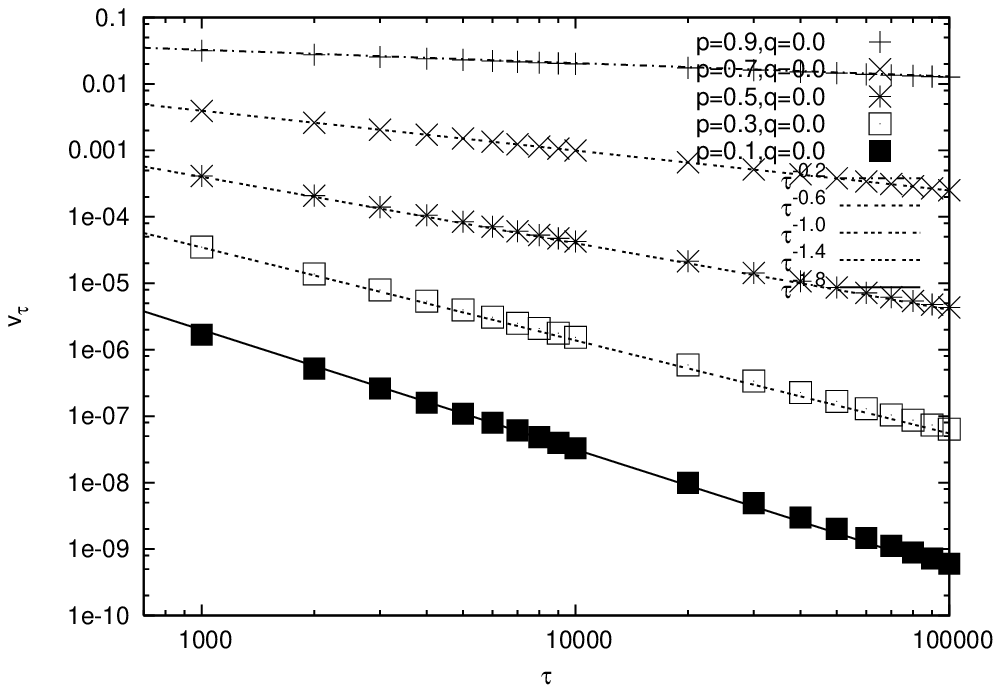}
\label{2}
\caption{Asymptotic behaviour of the scaled variance $\tilde{v}_\tau$ 
for $q=0$ and
$p=0.9,0.7,0.5,0.3,0.1$.}
\end{minipage}
\hspace{0.3cm}
\end{figure}
For  other  cases ($q=0,1$),
the slope changes continuously (Figure. 4).

\section{Concluding Remarks}

We investigated a voting model  that is similar to 
a Keynesian beauty contest.
Mathematically, our model is 
a binomial distribution (independent voters)  doped in a 
beta binomial distribution (copycat voters).
We calculated the exact solution for  special cases and
analyzed the general case using a stochastic differential equation.
In the special  cases, there is no phase transition. 
We will extend this function to the general $q$ in the future.
We believe that  the obtained  solution  is a  useful clue to 
understand  phase transitions clearly.

In  general, $q$, the correlation structure, exhibits a dramatic change 
at a critical value of the doping.
If copycats constitutes a  majority or number half of the total 
number of voters,
 the variance  converges  slower  than it would   in  a binomial distributions.
This implies    that our conclusion  is extremely volatile because 
the fundamental value becomes irrelevant, and it is difficult  to 
estimate the 
conclusion of the vote.

We  observed  phase transition in the limit $t\rightarrow \infty$. 
However, the long memory is  finite; therefore, this gives rise to 
the question of  whether  we  can  observe the  phase transition. 
This question arises because in this case,   voters are unable to 
see the     votes of all the  voters but can   only   see  the votes 
of   previous  voters. This model is  useful to understand the model 
introduced by Curty and Marsili.\cite{Curty}
We intend to address this issue in the future.

\section*{Acknowledgment}
This work was supported by
Grant-in-Aid for Challenging Exploratory Research 21654054 (SM).

\appendix
\def\thesection{Appendix \Alph{section}}
\section{}
We prove the assumption (\ref{rec}).
Multiplying (\ref{process})  
by $(-1)^{k-c_0} (l-1)!/\{(k-1)(k-2)\cdots c_0 (l-k+c_0-1)!\}=
(-1)^{k-c_0}(1)_{l-1}/{(c_0)_{k-c_0}(1)_{l-k+c_0-1}}$ 
and summing over $k=c_0, c_0+1,\cdots,l+c_0-1$ we get
\begin{eqnarray}
& &\sum_{k=c_0}^{l+c_0-1}(-1)^{k-c_0} 
\frac{(1)_{l-1}}{(c_0)_{k-c_0}(1)_{l-k+c_0-1}}P(k,t)\nonumber\\
&=&\sum_{k=c_0}^{l+c_0-1}(-1)^{k-c_0} 
\frac{(1)_{l-1}}{(c_0)_{k-c_0}(1)_{l-k+c_0-1}}P(k,t-1)-\frac{p}{c+t-2}
\nonumber \\
&\times&
\sum_{k=c_0}^{l+c_0-1}(-1)^{k-c_0} 
\frac{(1)_{l-1}}{(c_0)_{k-c_0} (1)_{l-k+c_0-1}}[kP(k,t-1)
-(k-1)P(k-1,t)]
],
\nonumber \\\
\label{6}
\end{eqnarray}
where 
$
(z)_i=z\cdot(z+1)\cdots(z+i-1).
$
We call the second and third terms of the  RHS  without 
the coefficient $p/(c_0+c_1+t-2)$ $A$ and  $B$, respectively.
We can rewrite $A$ as
\begin{eqnarray}
A&=& \sum_{k=c_0}^{l+c_0-2}(-1)^{k-c_0} 
\frac{(1)_{l-1}}{(c_0)_{k-c_0} (1)_{l-k+c_0-1}}kP(k,t-1)
\nonumber \\
& &
+(-1)^{l-1}\frac{(1)_{l-1}(l+c_0-1)}{(c_0)_{l-1} }P(l+c_0-1,t-1).
\end{eqnarray} 
We can rewrite $B$ as
\begin{eqnarray}
B&=&\sum_{k=c_0+1}^{l+c_0-1}(-1)^{k-c_0} 
\frac{(1)_{l-1}}{(c_0)_{k-c_0-1} (1)_{l-k+c_0-1}}P(k-1,t-1)\nonumber \\
&=&-\sum_{k=c_0}^{l+c_0-2}(-1)^{k-c_0} 
\frac{(1)_{l-1}(l-k+c_0-1)}{(c_0)_{k-c_0} (1)_{l-k+c_0-1}}P(k,t-1).
\end{eqnarray}
Then, $A-B$ is given by
\begin{eqnarray}
A-B&=&
(l+c_0-1)\sum_{k=c_0}^{l-1}(-1)^{k-c_0} 
\frac{(1)_{l-1}}{(c_0)_{k-c_0}(1)_{l-k+c_0-1}}P(k,t-1).
\label{9}
\end{eqnarray}
Substituting (\ref{9}) in (\ref{6}), we can obtain  the time evolution 
of the summation:
\begin{eqnarray}
& &\sum_{k=c_0}^{l-k-1}(-1)^{k-c_0} 
\frac{(1)_{l-1}}{(c_0)_{k-c_0}(1)_{l-k+c_0-1}}P(k,t)\nonumber\\
&=&
\frac{c+t-2-(l+c_0-1)p}{c+t-2}
\sum_{k=c_0}^{l-k-1}(-1)^{k-c_0} 
\frac{(1)_{l-1}}{(c_0)_{k-c_0} (1)_{l-k+c_0-1}}P(k,t-1).
\nonumber \\
\end{eqnarray}
For $k>C_0$, we can prove  that  the following equality holds:
\begin{eqnarray}
\sum_{k=c_0}^{l+c_0-1}(-1)^{k-c_0} 
\frac{(1)_{l-1}}{(c_0)_{k-c_0} (1)_{l-k+c_0-1}}P(k,t)
=\frac{(c-(l+c_0-1)p)_{t-1}}{(c)_{t-1}}.
\label{rec}
\end{eqnarray}
The analytic form can be obtained by multiplying both sides with 
$(-1)^{l-1}(k'-1)(k'-2)\cdots s/[(k'-l-c_0+1)!(l-1)!]
=(-1)^{l-1}(c_0)_{k'-c_0-1}/[(1)_{k'-l-c-0+1}(1)_{l-1}]$ and summing  over  
$l=1,\cdots,k'-c_0+1$

\begin{eqnarray}
 P(k',t)
&=&
\sum_{l=1}^{k'-c_0+1}(-1)^{l-1}\frac{(c_0)_{k'-c_0}}{(1)_{k'-c_0-l+1}(1)_{l-1}}
\frac{(c-(l+c_0-1)p)_{t-1}}{(c)_{t-1}}.
\end{eqnarray}
\section{}

Replacing the analytical form (\ref{mu}), we can obtain
\begin{eqnarray}
& &\hat{\mu}_r(t)=\sum_{k'=c_0}^{c_0+t-1}(k')_{r}
\sum_{l=1}^{k'-c_0+1}(-1)^{l-1}\frac{(c_0)_{k'-c_0}}{(1)_{k'-l-c_0+1}(1)_{l-1}}
\frac{(c-(l+c_0-1)p)_{t-1}}{(c)_{t-1}}.
\nonumber \\
& &
\end{eqnarray}
Further,
\begin{equation}
\sum_{k'=c_0}^{t+c_0-1}\sum_{l=1}^{k'-c_0+1}=
\sum_{l=1}^{t}\sum_{k'=l+c_0-1}^{t+c_0-1}
\end{equation}
and
\begin{eqnarray}
\sum_{k'=l+c_0-1}^{t+c_0-1}(k)_{r}\frac{(c_0)_{k-c_0}}{(1)_{k'-l-c_0+1}(1)_{l-1}}
=
\frac{(c_0)_{r+t}}{(1)_{l-1}(1)_{t-l}(l+c_0+r-1)}.
\end{eqnarray}
 the quasi-momentum can  have the following form:
\begin{eqnarray}
& &\hat{\mu}_r (t)=
\sum_{l=1}^{t}(-1)^{l-1} \frac{(c_0)_{t+r}}{(1)_{l-1}(1)_{t-l}(l+c_0+r-1)}
\frac{(c-(l+c_0-1)p)_{t-1}}{(c)_{t-1}}.
\nonumber \\
& &
\end{eqnarray}

\section{}
We use  $\delta X_\tau=X_{\tau+\epsilon}-X_\tau$ and $\zeta_\tau$, a 
standard iid Gaussian sequence; our objective is to identify the 
drift $f_\tau$ and variance $g^2_\tau$  such that
\begin{equation}
\delta X_\tau=f_\tau(X_\tau)\epsilon
+\sqrt{\epsilon}g_\tau(X_\tau)\zeta_{\tau+\epsilon}.
\end{equation}
Given $X_\tau=x$, using the transition probabilities of $\Delta_n$, we get
\begin{equation}
\textrm{E}(\delta X_\tau)=
\epsilon \textrm{E}(\Delta_{[\tau/\epsilon]+1}-\Delta_{[\tau/\epsilon]})
=\epsilon(2p_{[\frac{l/\epsilon+c+\tau/\epsilon-1}{2}],\tau/\epsilon}-1)
=
\epsilon[(1-p)(1-2q)+\frac{px}{\tau+1}].
\end{equation}
Then, the drift term is $f_\tau(x)=(1-p)(1-2q)+px/(\tau+1)$.
Moreover, 
\begin{equation}
\sigma^2(\delta X_\tau)=\epsilon^2
[
1^2p_{[\frac{l/\epsilon+c+\tau/\epsilon-1}{2}],\tau/\epsilon}
+(-1)^2(1-p_{[\frac{l/\epsilon+c+\tau/\epsilon-1}{2}],\tau/\epsilon})]
=\epsilon^2,
\end{equation}
  such  that 
$g_{\epsilon,\tau}(x)=\sqrt{\epsilon}.$
In  the continuous limit, we can obtain the Fokker-Plank  diffusion 
equation for this process:
\begin{equation}
\frac{\partial P}{\partial \tau}
=\frac{1}{2}
\frac{\partial^2 P}{\partial x^2}-
\frac{p}{\tau+1}\frac{\partial (xP)}{\partial x}
-(1-p)(1-2q)\frac{\partial P}{\partial x}.
\end{equation}

\end{document}